\documentclass{ws-ijmpa}
\newlength{\listwidth}
\begin{document}

%\markboth{Authors' Names}
%{Instructions for Typing Manuscripts (Paper's Title)}

\catchline{}{}{}{}{}
%
%%%%%%%%%%%%%%%%%%%%%%%%%%%%%%%%%%%%%%%%%%%%%%%%%%%%%%%%%%%%%%%%%%%%

%\title{THE JUPITER ELECTRON SCATTERING PROGRAM AT JEFFERSON LAB\\}
\title{The JUPITER Electron Scattering Program at Jefferson Lab\\}

\author{\footnotesize Arie Bodek\footnote{Talk given by Arie Bodek at DPF2004 Conference, to be published in
International Journal of Modern Physics.}}

\address{Department of Physics and Astronomy, University of Rochester, 
River Campus\\
Rochester, NY 14627, USA}

\maketitle

\begin{abstract}
   
JUPITER (Jlab Unified Program to Investigate 
nuclear Targets and Electroproduction of Resonances) is a new
 collaboration between the Nuclear Physics electron scattering
 and High EnergyÊPhysics neutrino scattering 
 communities to investigate the structure of
 nucleons and nuclei with electron and neutrino Beams.
 The first phase of JUPITER is Hall C experiment E04-001 on 
 Inclusive Electron Scattering from Nuclear Targets. First data run
 of E04-001 is currently scheduled for January of 2005.
 
\end{abstract}

%\section{Intorduction}	%) A SECTION HEADING

Experiments E02-109 and E04-001 at Jefferson Laboratory 
have been approved to measure the longitudinal-transverse (L-T) separated 
structure
functions $F_2$ and $R = \sigma_L / \sigma_T$
from Deuterium and  nuclear targets in the resonance region.
It is the first stage
of an overal JUPITER program to investigate quark-hadron duality and
electromagnetic and weak structure of nucleons and 
nuclei~\cite{P04-001,eois}.
The targets proposed include Carbon, Quartz, Aluminum, Calcium, Steel
and Copper in conjunction with  baseline data on Hydrogen and Deuterium.  
The targets
proposed are (or closely resemble) nuclear
targets commonly used in neutrino  experiments
such as the low statistics bubble chamber experiments on
Hydrogen, Deuterium, Neon, and  Argon and high statistics
neutrino experiments on Scintillator (MINERvA, MiniBONNE, K2K and JPARC), Water 
(SuperK, JPARC), Steel (MINOS) and Argon (CNGS). In addition
to 
performing studies of quark-hadron duality in electron scattering
on 
nuclear targets for the separated structure functions, these
data 
will used as input vector form factors in a future analysis
of 
neutrino data in order to investigate quark hadron-duality
in the 
axial structure function of nucleons and nuclei. This will be
done in collaboration with the MINERvA neutrino experiment~\cite{eois}
to be run in the NUMI~\cite{numi} low energy neutrino beam at Fermilab.

An immediate impact of these new measurements
with nuclear targets will be the reduction in
uncertainties in neutrino oscillation parameters for current and
near term neutrino oscillations experiments such as K2K and MINOS.
The data are even more important for the
more precise next generation neutrino
oscillations experiments such as JPARC and NUMI Off-axis.

Measurements will be made in the quasielastic and nucleon resonance region ($1<W^2<4$ 
GeV$^2$)
spanning the four-momentum transfer range $0.1<Q^2<4.0$ (GeV/c)$^2$.
The 
%separation 
decomposition
of the inclusive electroproduction cross sections into 
longitudinal
and transverse strengths will accomplished by performing Rosenbluth 
separations
to extract the transverse structure function $F_1(x,Q^2)$, the 
longitudinal structure function
$F_L(x,Q^2)$, and the ratio $R={{\sigma_L} / {\sigma_T}}$.  

The analysis of the separated resonance region proton data from 
experiment
E94-110~\cite{e94110}, which was also performed in Hall~C, has recently 
been completed.
Figures~\ref{2xf1f294110} and ~\ref{rfl94110}   show
preliminary results of analysis
of data from Jlab experiment E94-110~\cite{e94110} on hydrogen
in the resonance region. Similar  
data with  deuterium and with nuclear targets will be taken
by JUPITER. 

The extension of the proton measurements to deuterium and
nuclear targets is approved and
will done in a simultaneous run by E02-109~\cite{e02109}, and E04-001
The JUPITER program represents the first
global survey of these fundamental separated quantities in the 
resonance
region for nucleons bound in nuclei.
%UPDATE
Data from all of the previous
experiments will be also incorporated into the analysis.

The analysis
machinery which was developed for E94-110 can be used with only very 
little modification.
The great care and time invested in developing the experimental 
requirements, systematic uncertainty
measurements, and analysis machinery will be of immediate benefit to 
the program.
The program also builds upon the experience
from 
previous lower statistics studies with (electrons)  of the 
nuclear-dependence of the separated
structure functions in the DIS region that  were done by SLAC 
experiment E140~\cite{dasu}, as well as
the very low $Q^2$ high $W^2$ 
studies (to test the
HERMES effect) that are done by JLab experiment 
E99-118~\cite{e99118}. In parallel, a group of experimenters led by
Steve Manly (Rochester) and Will Brooks (Jlab) will investigate final states on nuclear
targets in Hall B.

 \begin{minipage}{0.5\textwidth}
\epsfxsize=3.0in \epsfbox{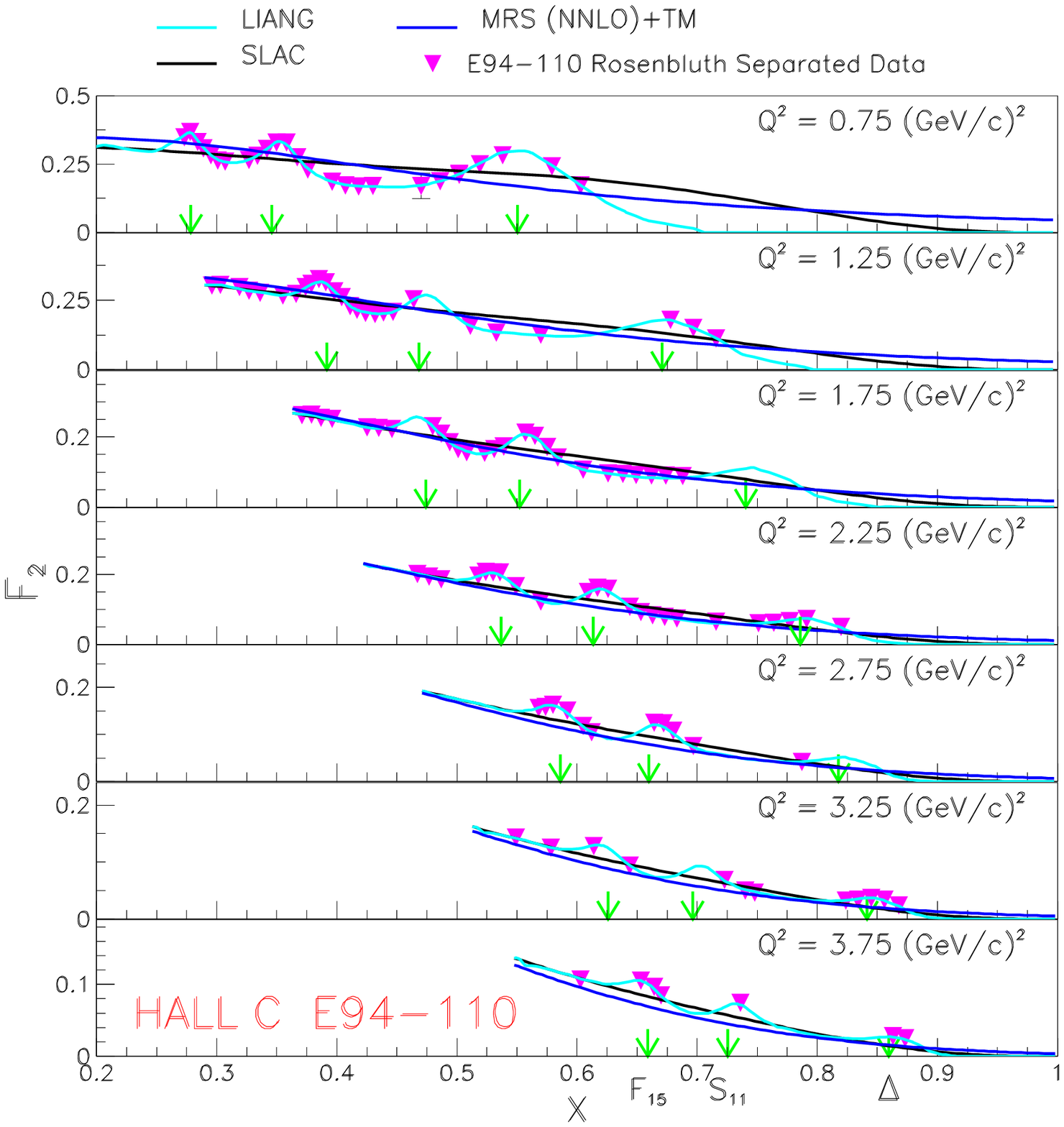}
\end{minipage}
    \begin{minipage}{\listwidth}
\epsfxsize=3.0in \epsfbox{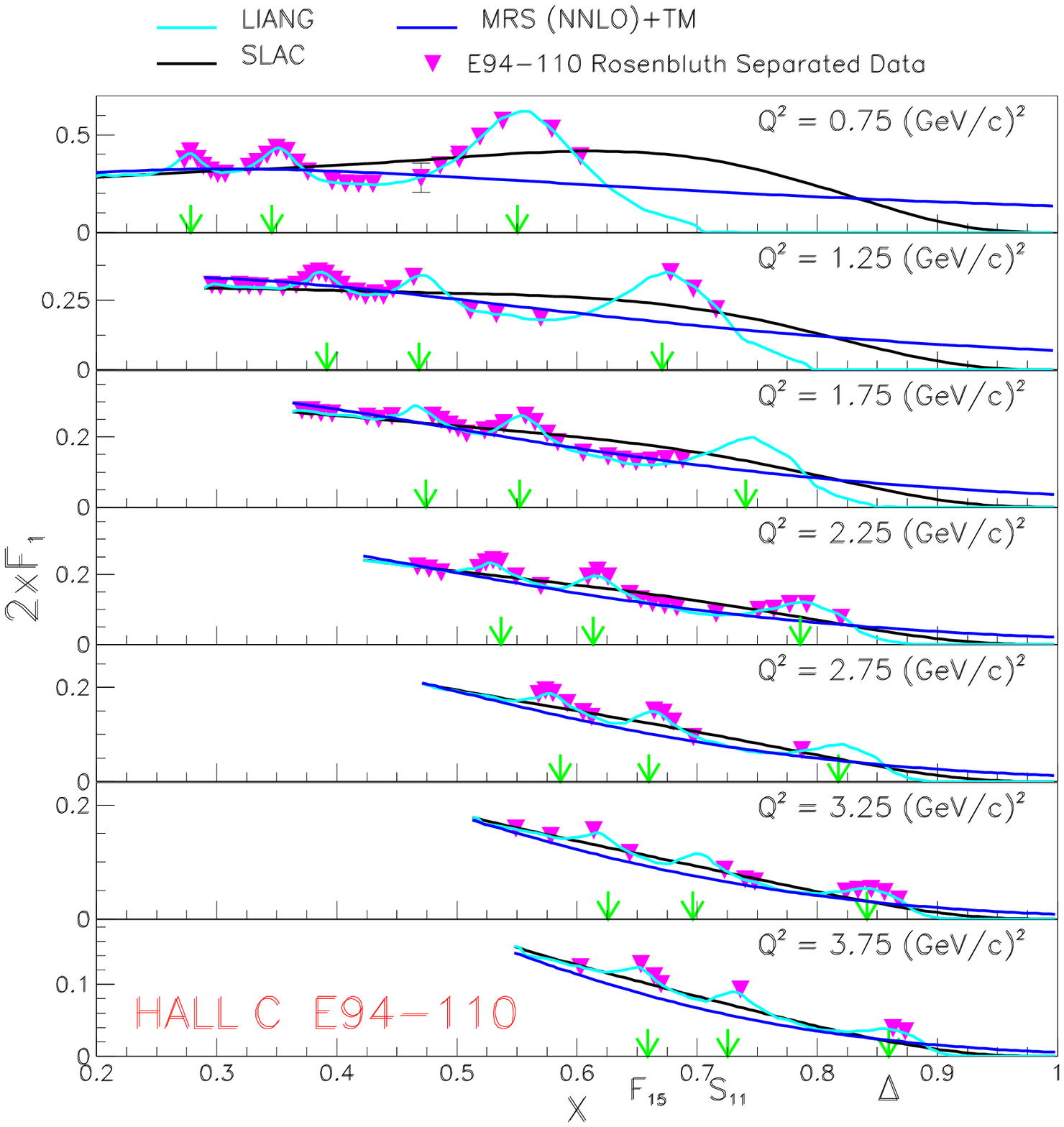}
\end{minipage}
\vspace{-0.2in}
\begin{figure}[h]
\caption{Recent data from Jlab experiment E94-110 
    (on Hydrogen) for $2xF_1$ (left) and $F_2$ (right)
    in the resonance region. JJUPITER Data with
    deuterium and nuclear targets
   ill be taken at the same time in a January 2005 run of E02-109/
  E04-001.}
\label{2xf1f294110}
\end{figure}

  \begin{minipage}{0.5\textwidth}
\epsfxsize=3.0in \epsfbox{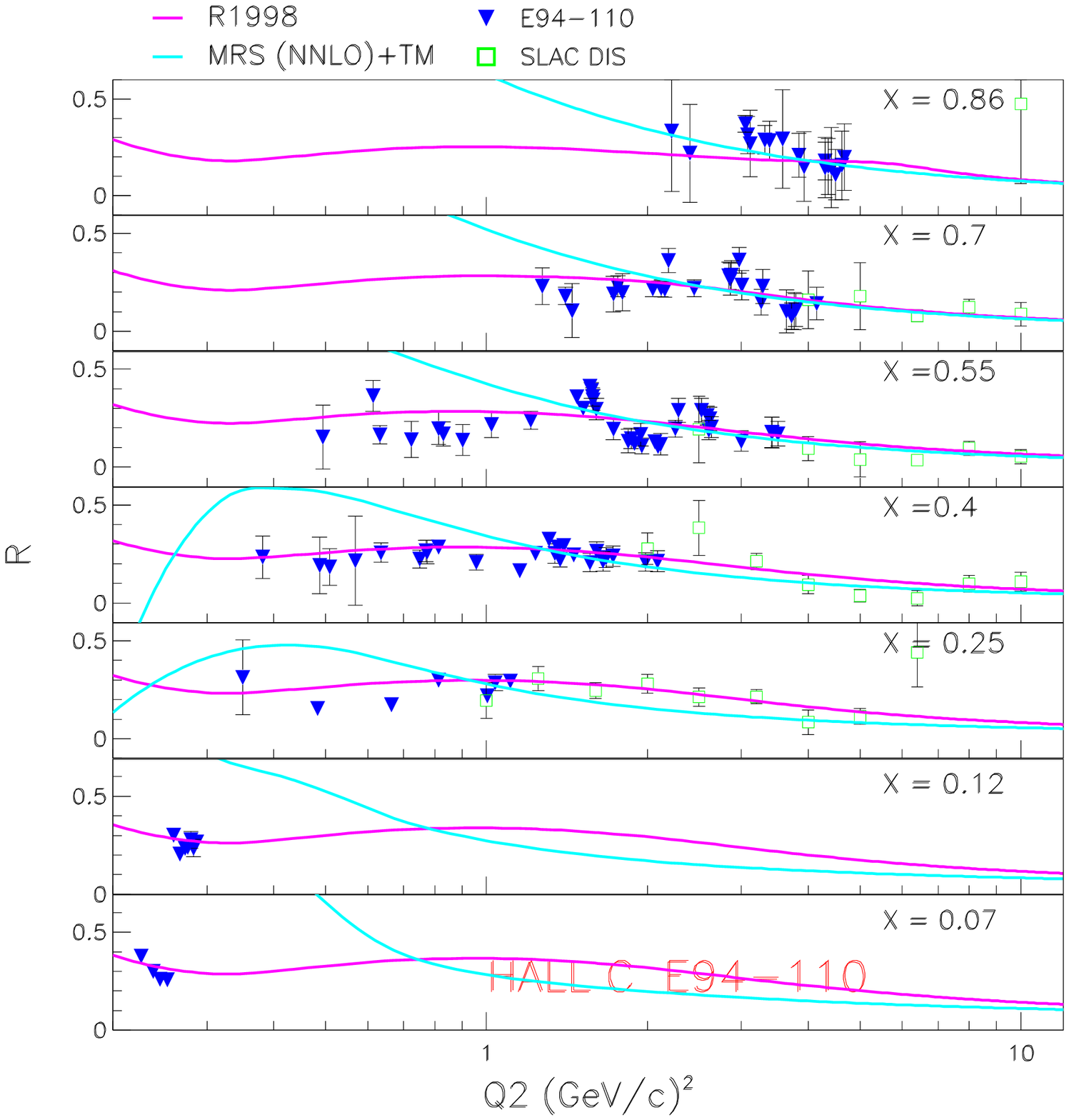}
\end{minipage}
    \begin{minipage}{\listwidth}
\epsfxsize=3.0in \epsfbox{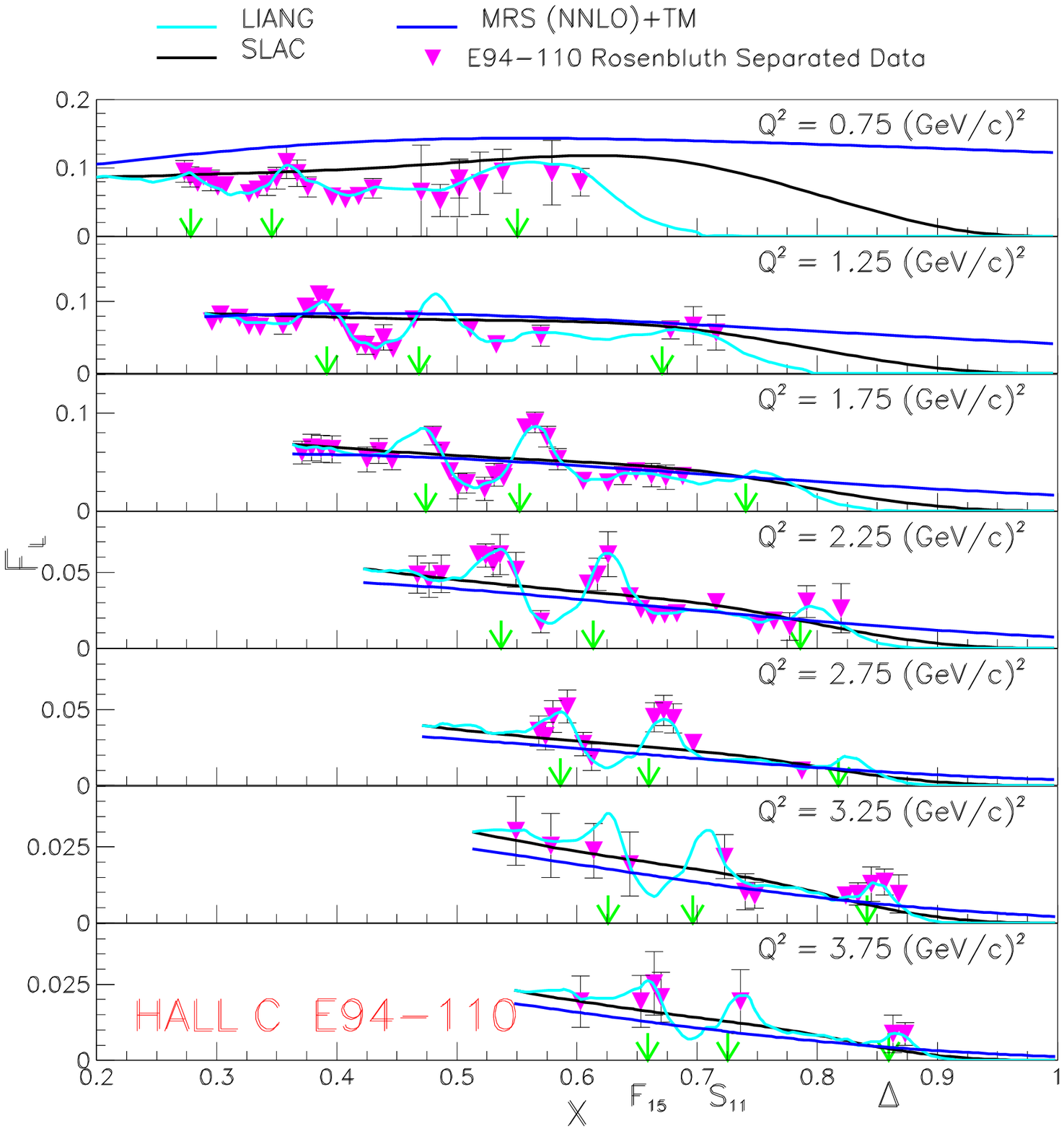}
\end{minipage}
\vspace{-0.2in}
\begin{figure}[h]
\caption{
Recent data from Jlab experiment E94-110 (on hydrogen) for $R$
(left) and $F_L$ (right)
    in the resonance region. JUPITER Data with
    deuterium and nuclear targets
   ill be taken at the same time in a January 2005 run of E02-109/
  E04-001.}
\label{rfl94110}
\end{figure}

\end{document}